\documentclass[12pt]{article}
\usepackage{a4wide}
\usepackage{graphicx}
\usepackage{amsmath}
\usepackage{slashed}

\begin{document}
{\renewcommand{\thefootnote}{\fnsymbol{footnote}}
\medskip
\begin{center}
{\LARGE A path-integral approach to the problem of time}\\
\vspace{1.5em}
M. M. Amaral\footnote{e-mail address: {\tt mramaciel@gmail.com}} and
Martin Bojowald\footnote{e-mail address: {\tt bojowald@gravity.psu.edu}}

\vspace{0.5em}
Institute for Gravitation and the Cosmos,\\
The Pennsylvania State
University,\\
104 Davey Lab, University Park, PA 16802, USA\\
\vspace{1.5em}
\end{center}
}

\setcounter{footnote}{0}

\begin{abstract}
  Quantum transition amplitudes are formulated for a model system with local
  internal time, using path integrals. The amplitudes are shown to be more
  regular near a turning point of internal time than could be expected based
  on existing canonical treatments. In particular, a successful transition
  through a turning point is provided in the model system, together with a new
  definition of such a transition in general terms. Some of the results rely
  on a fruitful relation between the problem of time and general Gribov
  problems.
\end{abstract}


 
\section{Introduction}

In relativistic systems, the lack of an absolute time parameter makes it
impossible to implement the usual requirement of unitary quantum evolution in
a straightforward manner. A local time variable is valid only for a finite
range, so that coordinate changes are required to patch together classical
trajectories.  Corresponding transformations between quantum theories,
describing the same system but based on different time choices, should then be
used in order to patch together piecewise quantum evolutions. However, if
evolution is unitary in each patch, it extends beyond the classical range of
the time parameter. The physical meaning of such an extension is unclear, in
particular if it happens in semiclassical regimes such as small curvature of a
cosmological model. Transformations between different time choices, even
global ones, are important also to guarantee covariance of the quantum
theory. Since such transformations have not been defined to complete
satisfaction, the problem of time \cite{KucharTime,IshamTime,AndersonTime}
remains open. (See \cite{ReducedKasner,MultChoice,ClocksDyn} for recent ideas
on transformations of time choices.) But before one can introduce such
transformations between different local internal times, the local quantum
evolution itself must be made well-defined, generalizing the usual unitary
evolution operators. An important question is then the behavior of evolution
near a turning point of local internal time, close to where it ceases to be a
good time variable.

Here, we begin a new investigation based on the relationship between time
choices and gauge fixings, both of which may be globally valid in simple
examples but in general hold only locally. The relationship between time and
gauge is easy to see because the absence of an absolute time in relativistic
systems is formally realized by having a Hamiltonian constraint, replacing the
usual Hamiltonian which can take different non-zero energy
values. Classically, the Hamiltonian constraint $C$ plays a dual role. It
generates evolution with respect to a fixed time choice by Hamiltonian
equations $\dot{f}=\{f,NC\}$, where different choices of the phase-space
function $N$ correspond to evolution with respect to different time
coordinates. The Hamiltonian constraint also generates transformations between
different time choices by the gauge transformations ${\rm d}f/{\rm d}\epsilon
= \{f,C\}$. A convenient way of fixing time choices, following Dirac
\cite{GenHamDyn1}, is to use an internal time, or to select one of the
phase-space variables, called $\phi$ in what follows, as the time parameter
$\tau$. For $\phi$ to be able to play the role of time at least locally, it
must be non-constant, so that $\dot{\phi}=\{\phi,NC\}\not=0$ in some part of
phase space. The time choice, expressed for instance as $G:=\phi-\tau=0$, then
amounts to a (local) gauge fixing of the Hamiltonian constraint.

The internal time as well as the gauge fixing are global if
$\{\phi,C\}=\{G,C\}$ does not become zero as $\tau$ changes on the family of
gauge-fixed hypersurfaces $C=0=G$. This condition is easily seen to be
realized in simple cases in which $C$ depends on the momentum $p_{\phi}$ of
$\phi$ via a standard kinetic term, but not on $\phi$ itself. We then have
$\{\phi,C\}\propto p_{\phi}$ and $\dot{p}_{\phi}=\{p_{\phi},C\}=0$. However,
most realistic field models have interaction potentials for all phase-space
variables, so that $\phi$-dependent potentials contribute to the Hamiltonian
constraint for all possible choices of $\phi$. In such cases, global internal
times do not exist, and gauge fixings of the form $\phi-\tau=0$ are subject to
Gribov problems \cite{Gribov}. (For a review of Gribov problems in gauge
theories, see for instance \cite{GribovRev}.)

Quantization in the presence of Gribov problems can be dealt with conveniently
if one uses path integrals. In the next section, we will therefore begin our
discussion by a brief review of these methods, making contact with previous
work on path-integral quantizations of free relativistic particles
\cite{PathIntFree,PathIntFree2}. Our new results are related to an extension
of these earlier constructions to model systems with time-dependent
potentials, which require local internal times. We will focus here on
non-technical issues in order to highlight interesting features in this new
situation. In Section~\ref{s:Turning}, we will suggest a possible definition
of evolution across turning points of local internal times, but also point out
open questions regarding a complete realization of this scheme. However, in
spite of these difficulties, the path-integral treatment is shown to have some
advantages compared with alternative approaches: it appears less singular
close to turning points than canonical descriptions which make use of
kinematical structures in intermediate derivations.

Local internal times have been implemented semiclassically in canonical
language, using methods of effective constraints
\cite{EffCons,EffTime,EffTimeLong}. A comparison with path-integral results is
instructive and is provided in Section~\ref{s:Compare}. In particular, the
canonical framework distinguishes between kinematical variables, some of which
take complex values, and physical ones on which reality conditions are
imposed. Path-integral versions show only the latter properties, which leads
to technical simplifications but also makes it more difficult to see some of
the underlying features, in particular those related to transformations of
time choices. Nevertheless, the successful evolution through a turning point,
provided here in a simple model system, is promising and suggests new
questions to be analyzed in the context of the problem of time.

\section{The problem of time as a Gribov problem}

We consider systems with two canonical pairs of degrees of freedom, $(q,p)$
and $(\phi,p_{\phi})$, subject to one constraint $C=0$ which for now will be
assumed to be of the form $C=p_{\phi}^2-H(q,p,\phi)^2$. (We will later specify
$H(q,p,\phi)$ in more detail.) Such systems have an action principle
\begin{equation}
  S=\int_{\tau_{\rm a}}^{\tau_{\rm b}}d\tau\left(p\dot{q}+p_{\phi}\dot{\phi}-NC\right)
\end{equation}
where $N$ is an auxiliary variable with zero momentum, $p_N=0$. Varying by $N$
imposes the constraint $C=0$.

Choosing $\phi$ as an internal time is equivalent to choosing the gauge fixing
$G=\phi-f(\tau)=0$ with a linear function $f(\tau)$. For now, we choose
$f(\tau)=\tau$. In order for this formulation to be meaningful, we have to
introduce a family of copies of the original phase space, labeled by the
parameter $\tau$. On each copy, $G=0$ sets $\phi$ to a constant value
$\tau$. The gauge fixing is global only if the gauge flow generated by the
constraint $C$ on $\phi$, $\delta\phi/\delta\epsilon= \{\phi,C\}$, has
solutions monotonic in $\epsilon$. Since the same equation can, in the
language of canonical general relativity, be interpreted as proper-time
evolution of $\phi$, a global gauge fixing is obtained if $\phi(\tau)$ evolves
without turning points. With our choice of $C$, turning points are reached
whenever the trajectory crosses the hyperplane $p_{\phi}=0$ in phase space.

One can see the relationship with $p_{\phi}=0$ in a more formal way by
considering the path integral of our gauge-fixed system. Combining the
constraint $C=0$ with the gauge-fixing condition $G=0$ turns our system into
one with a pair of second-class constraints. The constraint surface where
$C=0$ and $G=0$ then has a symplectic structure given by Dirac brackets
instead of the original Poisson brackets of canonical
variables. Correspondingly, the path-integral measure must be modified by a
factor which turns out to be equal to the Faddeev--Popov determinant
\cite{SecondClassPathInt,Anomaly}, or a simple Faddeev--Popov function
$|\{C,G\}|$ in the case of a single pair of constraints. For our general form
of $C$ and the chosen gauge fixing, $|\{C,G\}|=2|p_{\phi}|$ which vanishes
just at the turning points of $\phi$. The path-integral measure is degenerate
where $p_{\phi}=0$.

If $p_{\phi}=0$ is reached on a gauge orbit, the gauge fixing is not global
but subject to a Gribov problem. The hypersurface in phase space where
$p_{\phi}=0$ is called a Gribov horizon, separating regions in which the gauge
fixing may be used locally. In order to avoid overcounting degrees of freedom,
the path integration should be restricted to one of the Gribov regions, in
this case $p_{\phi}>0$ or $p_{\phi}<0$. This restriction can easily be
performed by inserting a step function $\theta(\pm p_{\phi})$ in the path
integral. We arrive at the path-integral expression
\begin{equation} \label{PathIntLong}
 (q_{\rm a}\phi_{\rm a}\tau_{\rm a}|q_{\rm b}\phi_{\rm b}\tau_{\rm b}) = \int
 DqDpD\phi Dp_{\phi}DN\cdot 2|p_{\phi}| \theta(\pm p_{\phi}) \delta(G)
 \exp\left(\frac{i}{\hbar} \smallint_{\tau_{\rm a}}^{\tau_{\rm
       b}}d\tau\left(p\dot{q}+p_{\phi}\dot{\phi}-NC\right)\right)
\end{equation}
for transition amplitudes, with paths restricted so that $q(\tau_{\rm
  a})=q_{\rm a}$, $q(\tau_{\rm b})=q_{\rm b}$, $\phi(\tau_{\rm a})=\phi_{\rm
  a}$ and $\phi(\tau_{\rm b})=\phi_{\rm b}$.

While the relationship between a measure factor of $2p_{\phi}$ in path-integral
treatments of the free relativistic particle and the Faddeev--Popov factor has
been recognized in \cite{PathIntFree}, the step function had in this paper
been inserted so as to restrict to positive frequencies, stating that it has
no counterpart in gauge theories. For a free particle, the relationship is
indeed not obvious because it has a constant $p_{\phi}$ on gauge trajectories,
so that the hypersurface $p_{\phi}=0$ is never crossed and does not constitute
a Gribov horizon. The language of Gribov problems in the context of internal
times is therefore meaningful only if the internal time used is local and has
a turning point. (A possible relationship between the Gribov problem and step
functions in path integrals for a free relativistic particle has been
conjectured but not pursued in \cite{PathIntFree2}.)

Choosing the negative sign of $p_{\phi}$, to be specific, we can solve the
constraint by $p_{\phi}=-|H(q,p,\phi)|$. This classical Hamiltonian also
appears in a path-integral version of (\ref{PathIntLong}) in which
integrations over $N$, $p_{\phi}$ and $\phi$ have been carried out explicitly:
Integration over $N$ in (\ref{PathIntLong}) leads to a delta function
\begin{equation}
 \delta(C)=\delta(p_{\phi}^2-H^2)= \frac{1}{2|p_{\phi}|}
 \left(\delta(p_{\phi}-|H|)+ \delta(p_{\phi}+|H|)\right)\,.
\end{equation}
The factor of $(2|p_{\phi}|)^{-1}$ cancels out with the Faddeev--Popov
function, while the first delta function is removed by the step function
restricting integrations to the Gribov region with negative $p_{\phi}$. A
single delta function is then left, which trivializes the integration over
$p_{\phi}$. The $\phi$-integration is trivialized by $\delta(G)$ for the gauge
fixing, setting $\phi=\tau$. Performing these integrations, we are left with a
standard path integral
\begin{equation} \label{Trans}
 (q_{\rm b}\tau_{\rm b}|q_{\rm a}\tau_{\rm a}) = \int DqDp
 \exp\left(i\hbar^{-1} \smallint_{\tau_{\rm a}}^{\tau_{\rm b}} 
 {\rm d}\tau (p\dot{q}- |H(q,p,\tau)|)\right)
\end{equation}
with a time-dependent Hamiltonian. At this stage, $\phi$ and $\tau$ are no
longer independent.

As long as no turning point of $\phi$ is crossed between $\tau_{\rm a}$ and
$\tau_{\rm b}$, this standard path integral can be used for our original
system with local internal time $\phi$. Evolution across a turning point is
more complicated. It requires us to change the branch from negative ${\rm
  d}\phi/{\rm d}\tau$ to positive ${\rm d}\phi/{\rm d}\tau$. We could use our
path integral written above to evolve all the way up to a turning point at
$\tau_{\rm t}$, continued with an analogous expression for the other branch to
go from $\tau_{\rm t}$ to the final $\tau_{\rm b}$. These two branches could
be connected by the usual composition rule
\begin{equation} \label{Prop}
 (q_{\rm b}\tau_{\rm b}|q_{\rm a}\tau_{\rm a}) = \int {\rm d}q_{\rm t}
 (q_{\rm b}\tau_{\rm b}|q_{\rm t}\tau_{\rm t})_+(q_{\rm t}\tau_{\rm t}|q_{\rm
   a}\tau_{\rm a})_-
\end{equation}
where the $\pm$-signs indicate the branch used.  At a turning point, the
original phase-space structure is ill-defined because the Dirac brackets on
the hypersurface $C=0=G$ are infinite, or the path-integral measure is
degenerate. However, after gauge fixing the path integrals in both branches,
$\phi$ and $p_{\phi}$ have been integrated out, so that the degeneracy of
their measure no longer plays a role.

Provided that the path integrals in both branches can be computed with a
time-dependent Hamiltonian $|H|$ near the turning point where $H=0$, evolution
through the turning point can be defined. A problem may arise at this stage
because $|H|$ is not regular at $H=0$ in many cases of interest. For a
time-dependent constraint in relativistic systems, for instance, the
Hamiltonian in (\ref{Trans}) may be of the form $H=\sqrt{p^2+m^2-V(\tau)}$
with a root-like pole at the turning point. Far from the turning point the
Hamiltonian can be approximated by using the ``non-relativistic'' expansion
$H=m+ (p^2-V(\tau))/2m+\cdots$, leading to standard path integrals. But close
to the turning point the square root cannot be expanded. These questions will
be shown to be solvable in a simple example discussed in the next section. The
main question, however, is how to implement a gauge fixing that takes into
account the fact that $\phi$ ``runs backwards'' after it reaches the turning
point. We will provide more details in the specific model of the next section.

\section{Evolution through a turning point}
\label{s:Turning}

The standard treatment of a global internal time $\phi$, as used in the
preceding section, makes use of a clear distinction between the two possible
signs of $p_{\phi}$. One of them, usually the negative sign so as to match up
with the common time dependence $\exp(-iEt/\hbar)$ of stationary states in
quantum mechanics, is taken as the one governing evolution forward in time,
while the opposite sign corresponds to backward evolution. If $\phi$ is a
local internal time with turning points, the clear distinction disappears
because on a single gauge orbit, $p_{\phi}$ changes sign. The variable to be
taken as internal time moves forward and backward on the same trajectory
$\phi(\epsilon)$, but it should also be possible to consider the trajectory in
reverse. There are now at least four different regimes --- forward and
backward evolution before as well as after the turning point --- and a choice
with only two options, such as the sign of $p_{\phi}$, is no longer sufficient
to distinguish all of them from one another.

In general, disentangling forward and backward evolution along trajectories
can be rather cumbersome, in particular when a local internal time with
several possible turning points is considered. We will therefore specialize
our model system further, so that we have a local internal time $\phi$ with a
single turning point. This feature can be achieved by choosing a linear
$\phi$-dependent potential, for which the constraint is
\begin{equation}
 C=p_{\phi}^2-p^2-m^2+\lambda\phi\,.
\end{equation}
Here, we also specialized the dependence on the evolving pair $(q,p)$. In
particular, there is no $q$-dependent potential (so that one could use $q$ as
a global internal time in this model) and $p$ is conserved. These properties
will imply further simplifications in the detailed construction. At the end of
our analysis we will briefly comment on more general models.

Before we proceed, we note that turning points and Gribov horizons can be
caused not only by potentials depending on internal time but also by
topological effects. An example for such topological turning points has
recently been discussed in detail in \cite{DiracChaos}. For instance, if the
configuration space of one coordinate $\phi$ is a circle, globally defined
basic variables are $\cos\phi$ and $\sin\phi$ together with $p_{\phi}$. A
simple-looking, $\phi$-independent constraint such as $C=p_{\phi}^2-p^2$ would
then lead to Gribov horizons on the non-trivial phase space because
$\{\cos\phi,C\}=-2p_{\phi}\sin\phi$ becomes zero even if the constant
$p_{\phi}$ is non-zero. However, such models and their turning points are
different from those studied here because they do not lead to the difficulty
of assigning signs of $p_{\phi}$ to different branches of orbits before and
after the turning point.

\subsection{Classical formulation}

While $\phi$ has a turning point at 
\begin{equation}
 \phi_{\rm t}= \frac{p^2+m^2}{\lambda}\,,
\end{equation}
$q$ is monotonic with respect to the gauge parameter $\epsilon$. We have 
\begin{equation} \label{qe}
 \frac{{\rm d}q}{{\rm d}\epsilon} = \{q,C\}=-2p
\end{equation}
and therefore $q(\epsilon)=q_0-2p\epsilon$ where $p$ is constant. In terms of
local $\phi$-evolution, we have
\begin{equation}
 \frac{{\rm d}q}{{\rm d}\phi} = \left(\frac{{\rm d}\phi}{{\rm
       d}\epsilon}\right)^{-1} \frac{{\rm d}q}{{\rm d}\epsilon}
 =-\frac{p}{p_{\phi}} \approx \pm\frac{p}{\sqrt{p^2+m^2-\lambda\phi}}
\end{equation}
for $p_{\phi}=\mp \sqrt{p^2+m^2-\lambda\phi}$. The solution is
\begin{equation} \label{qphi}
 q(\phi) = q_0\mp\frac{2}{\lambda}p\left(\sqrt{p^2+m^2-\lambda\phi}-
 \sqrt{p^2+m^2-\lambda\phi_0}\right)
\end{equation}
as long as $\phi$ and $\phi_0$ are connected by evolution that does not go
through the turning point of $\phi$. For $\phi$ moving toward the turning
point at $\phi_{\rm t}$, it is increasing so that
$|p_{\phi}|=\sqrt{p^2+m^2-\lambda\phi}$ decreases toward zero. By convention,
we then choose the negative sign for $p_{\phi}$ (the upper sign in
(\ref{qphi})), and $q(\phi)$ is increasing for positive $p$. In the other
branch, with $\phi$ decreasing at values smaller than $\phi_{\rm t}$, the
square root increases and $q(\phi)$ is growing if the opposite sign of
$p_{\phi}$ is used.  However, choosing the positive sign for $p_{\phi}$ (the
lower sign in (\ref{qphi})) should also mean that the original trajectory is
followed in reverse. This is the general problem of the lack of a clear
distinction between different branches of forward and backward evolution
before as well as after the turning point, mentioned in the beginning of this
section. We have to separate the choice of the sign of ${\rm d}\phi/{\rm
  d}\epsilon$ from the choice of the sign of $p_{\phi}$, even though they are
linked by ${\rm d}\phi/{\rm d}\epsilon=2p_{\phi}$ in classical equations of
motion. Such a separation turns out to be possible in path-integral
treatments.

Before we continue, it is useful to put our considerations in a more general
context, as realized for instance in models of general relativity. In such
cases, $\epsilon$ would be considered as proper time, the physical time
parameter used by observers. The success of any quantum prescription for
evolution should therefore be judged by comparison with
$\epsilon$-evolution. However, a parameter such as $\epsilon$ does not appear
in canonically quantized theories, thus motivating the use of an internal
time. If there is a monotonic relationship $\phi(\epsilon)$ for a global
internal time, $\phi$ and $\epsilon$ can be used interchangeably without
problems. If not, we have to find a way of describing evolution in terms of
$\phi$ in spite of its turning around. Such a description should be able to
include all branches of the classical $\epsilon$-evolution in which
semiclassical behavior is expected to be possible. (For instance, both
expansion and collapse should be realizable semiclassically in the
internal-time formulation of a recollapsing cosmological model.) Taking our
model as an example, we should be able to construct quantum evolution with a
monotonic $\langle\hat{q}\rangle(\tau)$ even across a turning point of
internal time, in correspondence with the monotonic $q(\epsilon)$ implied by
(\ref{qe}).

In order to solve this problem, we disentangle the branches by writing
$\phi$-evolution globally with a monotonic dependence on a time parameter
$\tau$. We introduce this new parameter so that $\phi=\tau$ if
$\lambda\tau<p_{\rm t}^2+m^2$ (before the turning point as measured by $\tau$)
and
\begin{equation} \label{phi2}
 \phi=2\phi_{\rm t}-\tau=2\lambda^{-1} (p_{\rm t}^2+m^2)-\tau
\end{equation}
if $\lambda\tau>p_{\rm t}^2+m^2$. The two ranges of $\tau$ correspond to the
two branches of $\phi$-evolution towards the turning point at $\phi_{\rm t}$
and away from it. The parameter $\tau$ defined in this way provides a
continuous and monotonic parameterization of the whole trajectory of
$\phi$. In particular, $q(\tau)$ is a monotonic function, in contrast to
$q(\phi)$. The extended range of $\tau$ ensures that we have now a clear
distinction of different phases, $p_{\phi}$ negative or positive for forward
and backward evolution, before the turning point if $\tau<\phi_{\rm t}$ and
after if $\tau>\phi_{\rm t}$. (At this point, a comparison with topological
turning points is useful: Even though the treatment of signs of $p_{\phi}$ is
then more straightforward, the extension of the time parameter given by
(\ref{phi2}) is similar to the unwinding of circular internal times
constructed in \cite{DiracChaos}.)

The ``time reflection'' introduced in the parameterization (\ref{phi2}) has
implications for the form of the $\phi$-Hamiltonian, governing evolution of
$(q,p)$ with respect to internal time $\phi$. In systems in which $\phi$ is a
global internal time, this Hamiltonian is just $p_{\phi}(q,p)$ obtained by
solving the constraint $C=0$ for $p_{\phi}$. In our model with a
$\phi$-dependent potential, the $\phi$-Hamiltonian is time dependent. For
evolution through the turning point, we should write this Hamiltonian in terms
of $\tau$ instead of $\phi$. Before the turning point we just replace $\phi$
in $p_{\phi}=-\sqrt{p^2+m^2-\lambda\phi}$ with $\tau$, so that $H_{\tau}(q,p) =
-\sqrt{p^2+m^2-\lambda\tau}$ if $\tau<\phi_{\rm t}$. After the turning point,
we replace $\phi$ by $\tau$ using (\ref{phi2}), and then have
\begin{equation} \label{Htauafter}
H_{\tau}(q,p)=-\sqrt{\lambda\tau-m^2-(2p_{\rm t}-p^2)}= -\sqrt{\lambda\tau-m^2-p^2}
\end{equation}
using the conservation of $p=p_{\rm t}$. The Hamiltonian
\begin{equation}
 H_{\tau}(q,p)= -\sqrt{|p^2+m^2-\lambda\tau|}=\left\{\begin{array}{cl}
     -\sqrt{p^2+m^2-\lambda\tau} & \mbox{ 
       if }\tau<\phi_{\rm t}\\ -\sqrt{\lambda\tau-m^2-p^2} & \mbox{
       if }\tau>\phi_{\rm t}\end{array}\right.
\end{equation}
generates the equation of motion
\begin{equation}
 \frac{{\rm d}q}{{\rm d}\tau} = \{q,H_{\tau}\} =-{\rm sgn}(p^2+m^2-\lambda\tau)
 \frac{p}{\sqrt{|p^2+m^2-\lambda\tau|}} \,.
\end{equation}
Even though we use the negative sign in solving for $p_{\phi}$ before and
after the turning point, the time reflection contained in $\phi(\tau)$ implies
that $q(\tau)$ after the turning point is described by backward evolution, In
order to have forward evolution, we should multiply the $\tau$-Hamiltonian
with ${\rm sgn}(p^2+m^2-\lambda\tau)$. In a path-integral treatment, to which
we turn now, this factor is provided automatically because the Hamiltonian for
the evolution of $(q,p)$ is derived from the term $\dot{\phi}p_{\phi}$ in the
action, not just $p_{\phi}$.

\subsection{Path-integral formulation}

We can transfer our classical parameterization to the path-integral
quantization if we change the original gauge-fixing condition for a global
internal time to
\begin{equation} \label{phiComp}
 \phi-\tau \theta(p_{\rm t}^2+m^2-\lambda\tau)- \left(\frac{2}{\lambda}
   (p_{\rm t}^2+m^2)-\tau\right) \theta(\lambda\tau-p_{\rm t}^2-m^2)=0\,,
\end{equation}
suitable for a local internal time with a single turning point.
Path-integrating over $\phi$ and $p_{\phi}$ solves the constraint and the
gauge-fixing condition, so that the $\phi$-dependence of the action is turned
into
\begin{equation}
 \dot{\phi}p_{\phi} = -\sqrt{p^2+m^2-\lambda\tau}\: \theta(p_{\rm
   t}^2+m^2-\lambda\tau)+ \sqrt{\lambda\tau-m^2-(2p_{\rm t}^2-p^2)}
 \:\theta(\lambda\tau-p_{\rm t}^2-m^2)\,.
\end{equation}
(Taking a $\tau$-derivative of the step functions in (\ref{phiComp})
contributes two delta functions, which however cancel out.) In the specific
model, $p=p_{\rm t}$ is conserved, so that the result can simply be written as
\begin{equation} \label{phidotp}
 \dot{\phi}p_{\phi} = -{\rm sgn}(p^2+m^2-\lambda\tau)
 \sqrt{|p^2+m^2-\lambda\tau|}\,.
\end{equation}
The Hamiltonian is therefore always real, but its sign changes according to
the branch of $\phi$-evolution (without changing the sign of $p_{\phi}$).

In this simple example with $q$-independent potential, the path-integral can
be computed in the momentum representation. We will first assume that only
ranges of evolution are considered which do not contain a turning point. Using
the general solution
\begin{equation}
 \psi(p,\phi)=c(p) \exp(2i(3\lambda\hbar)^{-1}(p^2+m^2-\lambda\phi)^{3/2})
\end{equation}
of the Schr\"odinger equation
\begin{equation}
i\hbar\frac{\partial\psi}{\partial\phi}=\sqrt{p^2+m^2-\lambda\phi}\:\psi
\end{equation}
obtained after quantizing the deparameterized constraint, it is easier to
compute the propagator directly instead of integrating over paths. Choosing a
complete set of functions
\begin{equation}
 c_x(p)=\frac{1}{\sqrt{2\pi\hbar}} e^{ixp/\hbar}\,,
\end{equation}
the propagator is given by
\begin{eqnarray} \label{PropLin}
 (p_{\rm b}\tau_{\rm b}|p_{\rm a}\tau_{\rm a}) &=&  \int {\rm d}x \psi_x(p_{\rm
   b},\tau_{\rm b}) \psi_x(p_{\rm a},\tau_{\rm a})^* \\
&=& \delta(p_{\rm b}-p_{\rm a})
 \exp\left(-2i(3\lambda\hbar)^{-1} \left((p_{\rm b}^2+m^2-\lambda\tau_{\rm
       b})^{3/2}- (p_{\rm a}^2+m^2-\lambda\tau_{\rm
       a})^{3/2}\right)\right)\,. \nonumber
\end{eqnarray}

The assumption that no turning point be contained in the range $(\tau_{\rm
  a},\tau_{\rm b})$ can be fulfilled only if the wave function is not
supported on momenta $p$ for which $p^2+m^2-\lambda\tau$ is negative for
$\tau$ in the given range. Any Gaussian clearly violates this assumption, so
that we have to be more careful with turning points even if we are interested
in semiclassical evolution close to a piece of a classical trajectory that
stays away from the turning point. We can, however, combine the specific
result (\ref{PropLin}) with (\ref{phidotp}) and the general composition rule
(\ref{Prop}), written in the momentum representation, in order to compute the
complete propagator for arbitrary initial states. We write
\begin{eqnarray}
 (p_{\rm b}\tau_{\rm b}|p_{\rm a}\tau_{\rm a}) &=& \int{\rm d}p_{\rm t}
 (p_{\rm b}\tau_{\rm b}|p_{\rm t}\tau_{\rm t})  (p_{\rm t}\tau_{\rm t}|p_{\rm
   a}\tau_{\rm a})\nonumber\\
&=& \int{\rm d}p_{\rm t} \left(\int DqDp \exp\left(\frac{i}{\hbar}
    \int_{\tau_{\rm t}}^{\tau_{\rm b}}
    (\dot{q}p+\sqrt{\lambda\tau-m^2-(2p_{\rm t}^2-p^2)})\right)\right)\nonumber\\
&&\times \left(\int DqDp \exp\left(\frac{i}{\hbar}
    \int_{\tau_{\rm a}}^{\tau_{\rm t}}
    (\dot{q}p-\sqrt{p^2+m^2-\lambda\tau})\right)\right)\nonumber\\
&=& \int{\rm d}p_{\rm t} \delta(p_{\rm b}-p_{\rm t})
\exp\left(2i(3\lambda\hbar)^{-1} (\lambda\tau_{\rm b}-m^2-(2p_{\rm t}^2-p_{\rm
    b}^2))^{3/2}\right)\nonumber\\
 &&\times \delta(p_{\rm t}-p_{\rm a})
\exp\left(-2i(3\lambda\hbar)^{-1} (p_{\rm a}^2+m^2-\lambda\tau_{\rm
    a})^{3/2}\right)\nonumber\\
&=& \delta(p_{\rm b}-p_{\rm a}) \exp\left( 2i(3\lambda\hbar)^{-1}
  \left((\lambda\tau_{\rm b}-m^2-p_{\rm a}^2)^{3/2}- (p_{\rm
      a}^2+m^2-\lambda\tau_{\rm a})^{3/2}\right)\right)\,.
\end{eqnarray}
No divergences related to the turning point are present.

We can now use the propagator to evolve an initial state ``through the turning
point.'' (Strictly speaking, such a state never evolves completely through the
turning point since it is supported on momenta $p$ for which, at any finite
$\tau$, $\phi$ has not yet reached its turning point.) We assume an initial
Gaussian state
\begin{equation}
 \psi(p_{\rm a},\tau_{\rm a})= \frac{1}{(2\pi)^{1/4}\sqrt{\sigma}}
   \exp\left(-\frac{(p_{\rm a}-p_0)^2}{4\sigma^2}- \frac{i}{\hbar} q_0p_{\rm
       a}\right)\,.
\end{equation}
It is easy to compute
\begin{eqnarray} \label{Gaussian}
 \psi(p_{\rm b},\tau_{\rm b}) &=& \int{\rm d}p_{\rm a} (p_{\rm b}\tau_{\rm
   b}|p_{\rm a}\tau_{\rm   a})  \psi(p_{\rm a},\tau_{\rm a})\nonumber\\
&=& \frac{1}{(2\pi)^{1/4}\sqrt{\sigma}}
   \exp\left(-\frac{(p_{\rm b}-p_0)^2}{4\sigma^2}- \frac{i}{\hbar} q_0p_{\rm
       b}\right)\nonumber\\
 &&\times \exp\left(2i(3\lambda\hbar)^{-1} \left(|p_{\rm
         b}^2+m^2-\lambda\tau_{\rm b}|^{3/2}- |p_{\rm b}^2+m^2-\lambda\tau_{\rm
         a}|^{3/2}\right)\right)\,.
\end{eqnarray}

The time-dependent expectation value of $\hat{q}$ in this state is given by
\begin{equation} 
\langle\hat{q}\rangle(\tau) \sim q_0- \frac{2}{\lambda}\left\langle \hat{p}
  \left(\sqrt{\hat{p}^2+m^2-\lambda\tau} - \sqrt{\hat{p}^2+m^2-\lambda\tau_{\rm 
      a}}\right)\right\rangle
\end{equation}
if $\tau$ and $\tau_{\rm a}$ are well before the classical turning point
belonging to $p_0$, and
\begin{equation} \label{qtau}
\langle\hat{q}\rangle(\tau) \sim q_0+ \frac{2}{\lambda}\left\langle \hat{p}
  \left(\sqrt{\lambda\tau-\hat{p}^2-m^2} + \sqrt{\hat{p}^2+m^2-\lambda\tau_{\rm 
      a}}\right)\right\rangle\,,
\end{equation}
if $\tau_{\rm a}$ is before and $\tau$ well after the turning point belonging
to $p_0$. (For $\tau$ not too close to the $p$-dependent turning point
$\tau_{\rm t}$, the spread of the Gaussian (\ref{Gaussian}) to values of $p$
for which $\tau<\tau_{\rm t}$ can be ignored in an approximation.) As required
for the classical behavior, discussed after Eq.~(\ref{qphi}),
$\langle\hat{q}\rangle(\tau)$ increases monotonically. Moreover, for a state
sharply peaked in the momentum, the expectation value in (\ref{qtau}) is well
approximated by the classical limit (\ref{qphi}), both before and after the
turning point. More precisely, for such a state one can approximate
\begin{eqnarray}
 \left\langle \hat{p} \sqrt{\lambda\tau-\hat{p}^2-m^2}\right\rangle &\approx& 
 \langle\hat{p}\rangle \sqrt{\lambda\tau-\langle\hat{p}\rangle^2-m^2}+
 \frac{{\rm d}^2\left(\langle\hat{p}\rangle
   \sqrt{\lambda\tau-\langle\hat{p}\rangle^2-m^2}\right)}{{\rm
     d}\langle\hat{p}\rangle^2} (\Delta p)^2\\
&=& \langle\hat{p}\rangle \sqrt{\lambda\tau-\langle\hat{p}\rangle^2-m^2}\left(1-
      \left(3-\frac{\langle\hat{p}\rangle^2}{\lambda\tau-
       \langle\hat{p}\rangle^2-m^2}\right) \frac{(\Delta
     p)^2}{\lambda\tau-\langle\hat{p}\rangle^2-m^2}\right) \nonumber 
\end{eqnarray}
up to higher moments of $p$. (We are using methods of canonical effective
equations, following \cite{EffAc,Karpacz}.) Even if the state remains starply
peaked in $p$ during evolution as the turning point is approached, the
semiclassical approximation to (\ref{qtau}) eventually becomes invalid close
to the turning point because of inverse powers of
$\lambda\tau-\langle\hat{p}\rangle^2-m^2\approx
\langle\hat{p}_{\phi}\rangle\approx 0$. This result is in agreement with
canonical effective treatments, which we will compare with in the next
section. But first, we briefly summarize general lessons that can be drawn
from our simple example.

\subsection{A formal definition of evolution through a turning point}

It is not straightforward to provide a good definition of evolution through a
turning point. As our example suggests, it is important to distinguish between
local relational evolution with respect to some internal time $\phi$, and
globally defined evolution with respect to a time parameter $\tau$.  The
latter should locally agree with the former (or its reverse) in small
intervals not including a turning point, so that it can be seen as patched-up
evolution obtained by matching different ranges of local evolution. Instead of
unitary evolution with respect to $\phi$, as required in the case of global
internal times, we can then impose two minimal conditions for meaningful
evolution through a turning point:
\begin{enumerate}
\item We have the composition law (\ref{Prop}) for transition amplitudes.
\item There should be semiclassical states well before and well after the
  turning point, whose quantum evolution is well approximated by the classical
  evolution.
\end{enumerate}
The first condition is a replacement of unitary evolution operators. The
second one is necessary because previous attempts to obtain evolution through
a turning point, based on deparameterization or the proposal of
\cite{WaldTime}, have led to a ``freezing'' of evolution after a turning point
\cite{WaldTimeModels,PhysEvolBI} which does not agree with classical
evolution. No freezing happens in our formalism, as seen in equation
(\ref{qtau}).

In our specific formulation of the second condition, we do not require
that a single state which is semiclassical before the turning point always
evolves to a semiclassical state after the turning point. However, using the
general transition amplitudes, it should be possible to follow an initial
state for a brief interval before or after the turning point, and obtain
nearly classical evolution for an appropriate choice of semiclassical
state. This property allows one to verify that different branches before and
after the turning point are indeed matched correctly, as we have seen in the
monotonic behavior of $\langle\hat{q}\rangle(\tau)$ in our example. (This
condition may have to be relaxed if there are intervals in which turning
points are reached in rapid succession. Such a behavior may leave no time to
set up semiclassical evolution \cite{EffTimeCosmo}. We could simply exclude
such ranges from the condition and consider the succession of turning points
as a single transition phase, on both sides of which semiclassical behavior
should be possible.)

\section{Comparison with canonical treatment}
\label{s:Compare}

It is difficult to reconcile the requirement of unitary evolution of wave
functions in a Hilbert space with the local nature of generic internal
times. We have arrived at an alternative of the canonical requirement of a
self-adjoint Hamiltonian or unitary evolution operator in the preceding
section. A consistent canonical formulation of local internal times can also
be obtained, at least semiclassically \cite{EffTime,EffTimeLong}, if one
describes states not by wave functions in a Hilbert space but as algebraic
expectation-value functionals. A state is then a positive linear functional
mapping the algebra of observables to the complex numbers. Conditions on
operators on wave functions in a Hilbert space are not as apparent in such a
formulation and may therefore be relaxed, although the price at which such a
generalization would be done is difficult to estimate at the semiclassical
level. Nevertheless, a comparison with the new results of a path-integral
formulation is of interest.

For constrained systems, there are two kinds of algebras of observables, the
kinematical one and the physical one. As shown by the treatment of effective
constraints, the constraints themselves as well as reality conditions can be
imposed by referring solely to expectation-value functionals and their
kinematical algebraic relationships, deriving in this way properties of
physical states without having to construct a physical Hilbert space with an
explicit inner product. The latter has so far been found only for systems with
global internal times, following \cite{Blyth}, so that it presents one of the
major obstacles toward a solution of the problem of time in canonical
terms. Also the problem of finding quantum observables is often a difficult
one. The algebraic methods of \cite{EffTime,EffTimeLong} map the quantum
problem to an analogous one of classical type, formulated on an enlarged phase
space. Classical methods and approximations can then be used to compute
expectation values and moments of quantum observables, without the need to
find operators for observables or a physical Hilbert space.

Canonical effective constraints are usually formulated for expectation-value
functionals parameterized by the expectation values and moments assigned by a
state to a set of basic variables. Using canonical basic operators $\hat{q}$
and $\hat{p}$, the variables are then $q:=\langle\hat{q}\rangle$ and
$p:=\langle\hat{p}\rangle$ as well as
\begin{equation}
 \Delta(q^ap^b) := \langle(\hat{q}-q)^a(\hat{p}-p)^b\rangle_{\rm Weyl}
\end{equation}
in totally symmetric or Weyl ordering. (For the sake of uniform notation, we
slightly modify the usual denotation of fluctuations by identifying
$\Delta(q^2)=(\Delta q)^2$.) A generalization to several canonical degrees of
freedom is straightforward. These variables form a phase space with Poisson
brackets derived from the commutator in the algebra, extending
\begin{equation} \label{Poisson}
 \{\langle\hat{A}\rangle,\langle\hat{B}\rangle\} :=
 \frac{\langle[\hat{A},\hat{B}]\rangle}{i\hbar}
\end{equation}
to moments by the Leibniz rule. The Hamiltonian flow generated by a function
on the space of states, with basic expectation values and moments as
coordinates, is then equivalent to the Schr\"odinger flow of the
expectation-value functional \cite{EffAc}.

Using the notation of our model system, $(q,p)$ along with $(\phi,p_{\phi})$
are kinematical variables. They are subject to a constraint $C=0$ and its
gauge flow. After quantization, physical states are those annihilated by the
constraint operator: $\hat{C}\psi=0$. For expectation-value functionals, this
means that their basic expectation values and moments are not arbitrary but
subject to several independent conditions
\begin{equation} \label{Cs}
 \langle\hat{C}\rangle=0\quad,\quad
 \langle(\hat{q}-q)\hat{C}\rangle=0\quad,\quad 
 \langle(\hat{p}-p)\hat{C}\rangle=0\quad,\quad
\langle(\hat{\phi}-\phi)\hat{C}\rangle=0\quad,\quad
 \langle(\hat{p}_{\phi}-p_{\phi})\hat{C}\rangle=0
\end{equation}
and so on with more factors of the basic kinematical variables on the left of
$\hat{C}$. (It is important to fix the ordering of $\hat{C}$ either to the
left or the right, so that a first-class constrained system is obtained
\cite{EffCons}.) All these conditions can be expressed as constraint functions
on expectation values and moments by using expansions such as
\begin{equation} \label{Cexpand}
 \langle C (\hat{q},\hat{p})\rangle = \langle
 C(q+(\hat{q}-q),p+(\hat{p}-p))\rangle = C(q,p)+ \sum_{a,b} \frac{1}{a!b!}
 \frac{\partial^{a+b}C(q,p)}{\partial^aq\partial^bp} \Delta(q^ap^b)
\end{equation}
and again straightforward extensions to several independent canonical pairs.
(The expansion is formal unless $\hat{C}$ is polynomial in basic operators.)
For our model constraint $C=p_{\phi}^2-H(q,p,\phi)^2$, we obtain
\begin{eqnarray}
 \langle\hat{C}\rangle &=& p_{\phi}^2- H(q,p,\phi)^2 +\Delta(p_{\phi}^2)\nonumber\\
&&-
 \left(\frac{\partial^2 H^2}{\partial q^2} \Delta(q^2)+
   \frac{\partial^2 H^2}{\partial p^2} \Delta(p^2)+ \frac{\partial^2
     H^2}{\partial 
     \phi^2} \Delta(\phi^2)\right.\nonumber\\
&&\left.+ 2\frac{\partial^2H^2}{\partial q\partial p}
   \Delta(qp)+ 2\frac{\partial^2H^2}{\partial q\partial \phi}
   \Delta(q\phi)+ 2\frac{\partial^2H^2}{\partial p\partial \phi}
   \Delta(p\phi)\right)\\
 \langle(\hat{q}-q)\hat{C}\rangle &=& 2p_{\phi} \Delta(qp_{\phi})- 2
 H(q,p,\phi) \left(\frac{\partial H}{\partial q} \Delta(q^2)+ \frac{\partial
     H}{\partial p}\Delta(qp)+ \frac{\partial H}{\partial \phi}
   \Delta(q\phi)\right)
\end{eqnarray}
and so on.

In our main example, we assume $H^2=p^2+m^2-\lambda \phi$, and have 
\begin{equation} \label{C}
 \langle\hat{C}\rangle = p_{\phi}^2- p^2-m^2+\lambda\phi+ \Delta(p_{\phi}^2)-
 \Delta(p^2)\,.
\end{equation}
Since we have expressed the quantum constrained system as a Hamiltonian one of
classical type, we can try to follow classical deparameterization techniques
as much as possible. Choosing $\phi$ as (local) internal time as before, we
should solve the constraint equation $\langle\hat{C}\rangle=0$ for the
momentum $p_{\phi}$. However, the quantum nature of our system implies that
not only expectation values $\phi$ and $p_{\phi}$ appear in the constraints,
but also moments such as $\Delta(p_{\phi}^2)$ in (\ref{C}). Solving (\ref{C})
for $p_{\phi}$ then does not give us a Hamiltonian for $\phi$-evolution of the
pair $(q,p)$ and their moments, unless we find a way to compute
$\Delta(p_{\phi}^2)$ independently.

This task is made possible by the presence of higher-order constraints in our
system (\ref{Cs}). In particular, the constraint
\begin{equation}\label{Cpphi}
\langle(\hat{p}_{\phi}-p_{\phi})\hat{C}\rangle= 2p_{\phi}\Delta(p_{\phi}^2)-
2p\Delta(pp_{\phi})+ \lambda(\Delta(\phi p_{\phi})-{\textstyle\frac{1}{2}}
i\hbar)=0
\end{equation}
provides a condition on $\Delta(p_{\phi}^2)$ independent of (\ref{C}). (There
is an imaginary term in (\ref{Cpphi}) because we are ordering $\hat{C}$ to the
right in (\ref{Cs}), while moments such as $\Delta(\phi p_{\phi})$ are ordered
totally symmetrically. Effective constraints are therefore complex-valued in
terms of kinematical variables. Reality conditions will be imposed below after
solving the constraints and eliminating their gauge flow, thereby
transitioning to physical states.)

Solving (\ref{Cpphi}) for $\Delta(p_{\phi}^2)$ still does not give us a
solution independent of $(\phi,p_{\phi})$-moments because there are
contributions from $\Delta(pp_{\phi})$ and $\Delta(\phi p_{\phi})$ in
(\ref{Cpphi}). We find an equation for $\Delta(pp_{\phi})$ if
we compute
\begin{equation} \label{Cp}
 \langle(\hat{p}-p)\hat{C}\rangle = 2p_{\phi}\Delta(pp_{\phi}) -2p\Delta(p^2)+
 \lambda \Delta(\phi p)\,,
\end{equation}
and one for $\Delta(\phi p_{\phi})$ given by
\begin{equation} \label{Cphi}
 \langle(\hat{\phi}-\phi)\hat{C}\rangle = 2p_{\phi} \Delta(\phi p_{\phi})+
 i\hbar p_{\phi} -2p\Delta(\phi p)+ \lambda \Delta(\phi^2)\,.
\end{equation}
The $(\phi,p_{\phi})$-moments seem to proliferate, but most of them are
subject to gauge flows generated by the constraints (\ref{Cs}). They can
therefore be eliminated when the gauge is fixed or factored out. As shown in
\cite{EffTime,EffTimeLong}, a good partial gauge fixing of the flow generated
by the constraints (\ref{Cpphi}), (\ref{Cp}), (\ref{Cphi}) and
$\langle(\hat{q}-q)\hat{C}\rangle$ (which latter we will not be using in
explicit form) is given by the conditions
\begin{equation} \label{Zeitgeist}
 \Delta(\phi^2)=\Delta(\phi q)=\Delta(\phi p)=0
\end{equation}
on moments of local internal time $\phi$, or the Zeitgeist according to
\cite{EffTime,EffTimeLong}.  (Three conditions suffice to fix the four
first-order constraints because the second-order moments do not form a
symplectic phase space, while (\ref{Poisson}) defines a non-invertible Poisson
bracket.) These gauge-fixing conditions implement the expectation that
choosing $\phi$ as internal time should turn $\phi$ into a time parameter of
classical type, which does not have non-trivial fluctuations or correlations.

The gauge-fixing conditions (\ref{Zeitgeist}) do not contain $\Delta(\phi
p_{\phi})$, which instead is determined by effective constraints. Using the
gauge-fixing conditions, we can solve (\ref{Cphi}) for
\begin{equation} \label{Deltappphi}
 \Delta(\phi p_{\phi})=-\frac{1}{2}i\hbar\,,
\end{equation}
(\ref{Cp}) for
\begin{equation}
 \Delta(pp_{\phi}) = \frac{p}{p_{\phi}} \Delta(p^2)\,,
\end{equation}
and (\ref{Cpphi}) for
\begin{equation}
 \Delta(p_{\phi}^2) = \frac{p}{p_{\phi}} \Delta(pp_{\phi})+
 \frac{i\hbar\lambda}{2p_{\phi}}= \frac{p^2}{p_{\phi}^2}\Delta(p^2)+
 \frac{i\hbar\lambda}{2p_{\phi}}\,.
\end{equation}
(Note that the imaginary value (\ref{Deltappphi}) ensures that the uncertainty
relation for $(\phi,p_{\phi})$ is formally obeyed even though
$\Delta(\phi^2)=0$ according to (\ref{Zeitgeist}).)  We finally obtain the
$\phi$-Hamiltonian for the evolution of $(q,p)$ and their moments by solving
\begin{equation} \label{Csol}
 \langle\hat{C}\rangle = p_{\phi}^2-p^2-m^2+ \frac{p^2-p_{\phi}^2}{p_{\phi}^2}
 \Delta(p^2)+ \lambda \phi+ \frac{1}{2} i\hbar \frac{\lambda}{p_{\phi}}=0\,.
\end{equation}

At this point, we have to select appropriate reality conditions. After solving
the constraints, the non-time variables $(q(\phi),p(\phi))$ and their moments
should become physical observables; they should therefore be real. Moreover,
$p_{\phi}$ will be used as the Hamiltonian generating the evolution of $(q,p)$
and their moments with respect to $\phi$, and therefore should be real
too. The remaining variables in (\ref{Csol}) cannot all be real because of the
imaginary term. The only consistent choice is to allow $\phi$ (defined so far
as the kinematical expectation value of $\hat{\phi}$) to take complex values,
since time is not an observable after deparameterization. The imaginary part
${\rm Im}\langle\hat{C}\rangle=0$ of (\ref{Csol}) then implies the imaginary
part
\begin{equation} \label{Imphi}
 {\rm Im}\phi= -\frac{\hbar}{2p_{\phi}}
\end{equation}
of time, while ${\rm Re}\phi$ remains unconstrained and free to play the role
of an evolution parameter.  The real part of (\ref{Csol}) is a quadratic
equation for $p_{\phi}^2$, giving the $\phi$-Hamiltonian
\begin{equation} \label{pphi1}
 p_{\phi} = \pm \sqrt{\frac{1}{2} \left(p^2+m^2-\lambda{\rm Re}\phi+
     \Delta(p^2)\pm 
 \sqrt{(p^2+m^2-\lambda{\rm Re}\phi+\Delta(p^2))^2- 4 p^2\Delta(p^2)}\right)}\,.
\end{equation}
We choose the positive sign for the interior square root, ensuring that we
have the classical Hamiltonian for small $\Delta(p^2)$. The remaining sign
choice is then the usual one, distinguishing forward evolution from backward
evolution. 

The Hamiltonian (\ref{pphi1}) is non-linear in the moment $\Delta(p^2)$. Since
we derived it from constraints expanded up to leading order in moments, we
should expand the square roots in (\ref{pphi1}) to the same order:
\begin{equation} \label{pphi}
 \pm p_{\phi} = \sqrt{p^2+m^2-\lambda{\rm Re}\phi}+ \frac{1}{2}
 \frac{m^2-\lambda{\rm Re}\phi}{(p^2+m^2-\lambda{\rm Re}\phi)^{3/2}}
 \Delta(p^2) = H(p,\phi)+
 \frac{1}{2} \frac{\partial^2H}{\partial^2p} \Delta(p^2)\,.
\end{equation}
The final expression is just the expansion of the deparameterized effective
Hamiltonian $\langle H(\hat{p},\phi)\rangle$ as in (\ref{Cexpand}) in which
$\phi$ is treated as internal time rather than an operator. This result agrees
with the path-integral expression (\ref{Trans}) obtained after gauge fixing.

One can solve for some of the moments as well. For fluctuations of $p_{\phi}$,
for instance, we obtain
\begin{equation}
 \Delta(p_{\phi}^2) = \frac{p^2}{p^2+m^2-\lambda{\rm Re}\phi} \Delta(p^2)+
 \frac{1}{2}i\hbar \frac{\lambda}{\sqrt{p^2+m^2-\lambda{\rm Re}\phi}}
\end{equation}
which is not real.  The expectation value of $\hat{p}_{\phi}^2$ is then
\begin{equation}
 \langle\hat{p}_{\phi}^2\rangle = \Delta(p_{\phi}^2)+p_{\phi}^2=
 p^2+m^2+\Delta(p^2)- \lambda \left({\rm Re}\phi-\frac{1}{2}i \hbar
   \frac{1}{\sqrt{p^2+m^2-\lambda\phi}}\right)\,.
\end{equation}
This is the correct result obtained from $\langle\hat{C}\rangle=0$, taking
into account an imaginary part of $\langle\hat{\phi}\rangle$ (and treating
$\hbar\Delta(p^2)$ as a higher-order term).

If we could use a physical Hilbert space while treating $\phi$ as an internal
time, only $\hat{q}$ and $\hat{p}$ would be well-defined among the basic
operators. While expectation values and moments of $(\phi,p_{\phi})$ exist on
the kinematical Hilbert space and are real, they do not exist as independent
variables on the physical Hilbert space, and they need not be subject to
reality conditions. Only the evolution generator in internal time, given by
$\langle\hat{p}_{\phi}\rangle$ as a function of $(q,p)$-moments, must be real
as used above. The resulting imaginary part of the kinematical
$\langle\hat{\phi}\rangle$ turns out to be an important property of solutions
to effective constraints, because it allows one to change local internal times
by gauge transformations. If one transforms from internal time $\phi$ to
internal time $q$, for instance, the imaginary part of $\phi$ is removed and
replaced by an imaginary part of the new time $q$. (For details on this
result, which are rather technical, we refer to \cite{EffTime,EffTimeLong}.)

If transformations between local internal times can be implemented at the
level of path integrals, an imaginary part of the time expectation value would
seem to be an important property as well. However, after gauge fixing, it is
no longer possible to compute a time expectation value. In fact, the path
integral of a constrained system, which contains a factor of $\delta(C)$ in
its integrand, amounts to a projector on the physical Hilbert space on which
$\phi$ does not exist as an operator. It is conceivable that an imaginary
contribution to time appears in two possible ways: (i) One could impose a
gauge fixing which mixes different time choices, such as $G=\epsilon
q+(1-\epsilon)\phi-\tau=0$ for fixed $\tau$, with a transition of $\epsilon$
from zero to one switching time. With such a choice, factors from the
Faddeev--Popov determinant and the delta function $\delta(C)$ no longer cancel
out, so that the relevant path integrals are more complicated. (ii) A careful
discretization of paths in the unfixed integral with a time-dependent
constraint could show subtleties that require an imaginary contribution. We
will not pursue these rather technical issues here, but note the main result
of this section: A comparison of evolution (\ref{pphi}) in the canonical
effective treatment (to leading semiclassical order) agrees with what one
would expect from the path-integral formula (\ref{Trans}), before and after a
turning point.

\section{Implications}

In physical terms, the problem of time has two main aspects, both related to
the choice of internal time variables for relational evolution. For the most
part, we have addressed the question of how one can define evolution through a
turning point of a local internal time, and only briefly commented on the
possibility of changing the choice of time. The former aspect is important for
a complete definition of relational evolution, while the latter is crucial if
one tries to ensure covariance of the quantum system in the sense that it
provides predictions independent of the choice of internal time. Our
path-integral treatment has, so far, not led to new results on the question of
transforming between different internal times, for which kinematical aspects
and non-real time expectation values seem to play a role according to the
canonical treatment of \cite{EffTime,EffTimeLong}.

Regarding evolution through a turning point of a local internal time, we have
provided a specific definition based on an example solved explicitly. The
path-integral treatment suggests several simplifications compared with the
canonical one. Solutions of effective constraints lead to quantities which are
singular at the turning point where $p_{\phi}=0$, seen for instance in the
imaginary part of time (\ref{Imphi}) or in semiclassical corrections to the
$\phi$-Hamiltonian in (\ref{pphi}). However, the imaginary part of time is a
kinematical quantity and therefore auxiliary, and the contribution of
(\ref{pphi}) singular at $p_{\phi}=0$ has been obtained from an expansion of
(\ref{pphi1}) which does not appear to be singular at the turning point. The
effective treatment, to orders considered so far, cannot tell whether the
singularity in (\ref{pphi}) is only apparent or real. In
\cite{EffTime,EffTimeLong}, turning points of internal times therefore had to
be evaded by transforming to a different internal time before a turning point
is reached. While such transformations are possible in the effective
treatment, they can, as pointed out in \cite{EffTimeCosmo}, lead to problems
in regimes in which turning points of different variables are close to one
another, such as chaotic systems. Such transformations may be difficult to
define at the path-integral level, but our results here make one promising
suggestion: Turning points appear to be much less singular in this formalism,
so that we can evolve right up to the turning point from one side and, after a
``reflection'' (\ref{phi2}) of time, continue onwards. The evolved state then
indeed suggests that a semiclassical treatment should lead to difficulties
near a turning point of local internal time, as seen explicitly in the
expansion of $\langle\hat{q}\rangle(\tau)$ by moments in our model: There are
not only terms of the standard semiclassical form $\Delta(p^2)/p^2$, which can
always be chosen small for a suitable semiclassical state, but also
$\Delta(p^2)/p_{\phi}^2$, which are large near $p_{\phi}=0$ for any
state. Without the moment expansion, however, we have achieved a direct
matching of branches before and after the turning point, without any
divergence or freezing (as in \cite{WaldTimeModels,PhysEvolBI}) of the
evolution. Such a direct matching might eliminate problems encountered in
semiclassical treatments of chaotic systems, pointed out in
\cite{EffTimeCosmo}.

In our model system, a well-defined quantum evolution through a turning point
of internal time has been provided, but the proposal is not free of potential
practical problems in general. For instance, the reflection (\ref{phi2})
refers to the values which the evolving variables take at the turning
point. In our simple example, this expression depends only on the momentum $p$
which is a constant of motion. The value $p_{\rm t}$ at the turning point
could therefore be identified with the momentum $p$ at any other time. With
this substitution one obtains a $\phi$-Hamiltonian after the turning point
which depends only on the evolving canonical variables $(q,p)$. More
generally, there would be a complicated relation between $(q_{\rm t},p_{\rm
  t})$ and the canonical variables $(q,p)$ at some other time, or the
$\phi$-Hamiltonian after a turning point would be non-local in time if
$(q_{\rm t},p_{\rm t})$ are left without expressing them in terms of the
evolving pair $(q,p)$. (Even the corresponding classical evolution governed by
(\ref{Htauafter}) would be non-local in $\phi$-time.)

Our analysis has led us to a new treatment of the sign of $p_{\phi}$ (the
momentum of internal time) which differs from the usual choices for global
internal times. In a path integral, the Hamiltonian for evolution of the
non-time variables $(q,p)$ ends up being $\dot{\phi}p_{\phi}$ rather than just
$p_{\phi}$. In order to disentangle forward and backward motion of a local
internal time $\phi$ into a global time parameter $\tau$, we were led to
choosing a gauge fixing (\ref{phiComp}) that includes a ``time reflection''
after the turning point. The time parameter $\tau$ can then continue running
forward, even while the phase-space variable $\phi$ moves back after its
turning point. The time reflection changes the sign of $\dot{\phi}$, making it
unnecessary to switch to a different sign of $p_{\phi}$. At this point, the
analogy of the problem of time with general Gribov problems provides a
justification for the new procedure, because the whole evolution can be
formulated within one Gribov region. It is questionable whether this behavior
can be modeled in a canonical treatment, where the Hamiltonian for
$\phi$-evolution is $p_{\phi}$ rather than $\dot{\phi}p_{\phi}$. This
difference between canonical and path-integral treatments may be the reason
why it has been difficult to provide meaningful canonical evolutions through
a turning point of local internal time.

\section*{Acknowledgements}

We are grateful to Philipp H\"ohn for discussions. 
M. M. Amaral acknowledges support from CNPq, Brazil.  This work was supported
in part by NSF grant PHY-1307408.


\end{document}